# Graphical model approximations of random finite set filters*

Jason L. Williams
Intelligence, Surveillance and Reconnaissance Division
Defence Science and Technology Organisation, Australia
email jason.williams@dsto.defence.gov.au

17 August 2011

**Abstract**

Random finite sets (RFSs) has been a fruitful area of research in recent years, yielding new approximate filters such as the probability hypothesis density (PHD), cardinalised PHD (CPHD), and multiple target multi-Bernoulli (MeMBer). These new methods have largely been based on approximations that side-step the need for measurement-to-track association. Comparably, RFS methods that incorporate data association, such as Morelande and Challa's (M-C) method, have received little attention. This paper provides a RFS algorithm that incorporates data association similarly to the M-C method, but retains computational tractability via a recently developed approximation of marginal association weights. We describe an efficient method for resolving the track coalescence phenomenon which is problematic for joint probabilistic data association (JPDA) and related methods (including M-C). The method utilises a network flow optimisation, and thus is tractable for large numbers of targets. Finally, our derivation also shows that it is natural for the multi-target density to incorporate both a Poisson point process (PPP) component (representing targets that have never been detected) and a multi-Bernoulli component (representing targets under track). We describe a method of recycling, in which tracks with a low probability existence are transferred from the multi-Bernoulli component to the PPP component, effectively yielding a hybrid of M-C and PHD.

## 1 Introduction

Recently much work has been devoted to RFS-based approximations such as the [C]PHD and MeMBer filters [1]. A key characteristic of these methods has been their tractability, which is the consequence of clever approximations that

*This paper is UNCLASSIFIED and is approved for public release.

avoid data association. Algorithms such as M-C [2] have received little attention due to the computational cost associated with calculating marginal association probabilities, *i.e.*, similar to those used in joint probabilistic data association (JPDA) [3]. This paper develops a tractable version of the M-C algorithm based on a recently-developed, high quality, tractable approximation to these marginal probabilities [4,5]. Repeating the derivation of M-C, we obtain a form that permits application of this approximation, and show that it is natural for the multiple target distribution to contain both a component representing targets that have been detected, and a PPP distribution representing targets that so-far remain undetected, as previously suggested in [6].

Section 2 provides this derivation of prediction and update steps for the full Bayes multi-target distribution. Naturally this is intractable; difficulties in implementation occur in four areas:

1. Dealing with the exponential growth of hypotheses within each track
2. Appropriately representing the PPP distribution of undetected targets
3. Dealing with the growth in the number of tracks as new measurements are received
4. Performing inference on the joint target distribution (*i.e.*, data association)

The first difficulty has been well-examined in previous works such as [7] and [8]. In this paper, we argue that the second can be addressed via a grid filter using a relatively coarse discretisation (since the primary motivation is to represent the intensity of targets that have never been detected). In Section 6, we show that this solution can then aid in resolution of the third difficulty through *recycling*, which represents tentative tracks (those with low probability of existence) via the PPP distribution. Finally, we show in Section 3 that our previous results on LBP data association in [4,5] can be leveraged to address the fourth difficulty. We refer to the resulting algorithm as the Belief Propagation Marginal Track Filter (BPMTF).

The BPMTF is effectively a hybrid of a PHD representing low confidence targets, and a MeMBer distribution representing high confidence targets. Our derivation also admits an alternative algorithm that maintains several scans of history, and thus the prior for the existing tracks at each scan is not multi-Bernoulli. Again we rely on BP to calculate marginal association weights; in this case, the weights calculated are for association *history* events, rather than just association events in the latest scan, similarly to multiple scan JPDA [9]. We refer to this as the multiple scan BP filter (MSBPF). In contrast to multiple dimensional assignment (MDA) [10], which calculates a multiple scan MAP association, the MSBPF approximately calculates multi-scan marginal probabilities. While maintaining inter-target dependence in the association history is essential in hard association systems (*e.g.*, MDA), typically less benefit is observed in soft association systems such as the one studied here (*i.e.*, systems which calculate marginal weights). Therefore, we expect that the performance gain of the MSBPF over the BPMTF may not be large, such that the lower complexity BPMTF may ultimately be the preferred method. Some initial results of this comparison were provided in [4]; details of the MSBPF and further experiments will be the subject of a later publication.

Finally, in Section 4 we describe a method for addressing the problem of coalescence, which affects most algorithms that calculate marginal association weights, including JPDA and M-C. The proposed algorithm is based on efficient network flow optimisation methods [11], and thus is tractable for problems involving large numbers of targets.

In our development we neglect implementation issues such as gating, clustering and mixture reduction. Any practical implementation would obviously include these concepts; we leave these details to the reader.

# 2 Random set filter derivation

The derivation that follows largely parallels [2], making extensive use of models, methods and results in [1]. The assumptions in our development include:

- Targets arrive at time $t$ according to a non-homogeneous PPP with intensity $\lambda^{\mathrm{b}}(x)$ (let $\lambda^{\mathrm{b}} \triangleq \int \lambda^{\mathrm{b}}(x)\mathrm{d}x$), independent of existing targets
- Targets depart according to independent, identically distributed (iid) Markovian processes; the survival probability in state $x$ at time $t$ is $P^{\mathrm{s}}(x)$
- Motion for each target is governed by a Markovian process, independent of all other targets; the single-target transition probability density function (PDF) is $f_{t+1|t}(x|x')$
- Each target may give rise to at most one measurement; the probability of detection in state $x$ at time $t$ is $P^{\mathrm{d}}(x)$
- Each measurement is the result of at most one target
- False alarms arrive at time $t$ according to a non-homogeneous PPP with intensity $\lambda^{\mathrm{fa}}(z)$, independent of targets and target-related measurements
- Each target-derived measurement is independent of all other targets and measurements conditioned its corresponding target; the single target measurement likelihood is $f(z|x)$

We denote by $Z_t = \{z^1, \dots, z^{m_t}\}$ the measurement set at time $t$, and by $Z^t = \{Z_1, \dots, Z_t\}$ the measurement history up to and including time $t$. The development does not change if the birth density, survival probability, detection probability, false alarm density and measurement likelihood are time varying (*e.g.*, depending on a known, time-varying sensor state); we omit the time index from these parameters for notational simplicity.

The form which we assume for the multiple target probability distribution at time $t$ incorporates the following elements:

- A PPP intensity of undetected targets, $\lambda^u_{t|t}(x)$
- A set of resources[1] $\mathcal{R}_t$, the elements of which are of the form $(t, j)$, where $j \in \{1, \ldots, m_t\}$ is an index of a measurement in scan $t$
- A series of Bernoulli tracks, $i \in \mathcal{T}_t = \{1, \ldots, n_t\}$, each incorporating:
  - A set of hypotheses $\mathcal{H}^i_t$, the elements of which denote resource consumption[1] (for each $a \in \mathcal{H}^i_t$ we will have $a \subseteq \mathcal{R}_t$)
  - For each $a \in \mathcal{H}^i_t$, a hypothesis weight $w^{i,a}_{t|t}$
  - For each $a \in \mathcal{H}^i_t$, a hypothesis-conditioned Bernoulli multi-target distribution[2]

$$f^{i,a}_{t|t}(X) = \begin{cases} 1 - q^{i,a}_{t|t}, & X = \emptyset \\ q^{i,a}_{t|t} f^{i,a}_{t|t}(x), & X = \{x\} \\ 0, & |X| \geq 2 \end{cases}$$

where $q^{i,a}_{t|t}$ is the probability of existence under the hypothesis, and $f^{i,a}_{t|t}(x)$ is the existence-conditioned PDF

Following [4,5], we also define resource compatibility tables $\psi^{i,j}_r(a, b) \in \{0, 1\}$ for all $i \in \mathcal{T}_t$, $j \in \mathcal{R}_t$, $a \in \mathcal{H}^i_t$, $b \in \mathcal{T}_t$ as:

$$\psi^{i,j}_r(a, b) = \begin{cases} 0, & b = i,\ j \notin a \text{ or } b \neq i,\ j \in a \\ 1, & \text{otherwise} \end{cases} \tag{1}$$

These functions are used within the graphical models formalism to encode the constraints that measurements can be used at most once in any association event (see Sections 2.2 and 3, and [4,5]). Loosely, they enforce global consistency by maintaining two redundant representations of the hypothesis (variables $a$ comprised of elements indicating the association hypothesis for each target, and variables $b$ comprised of elements indicating the association hypothesis for each measurement), and ensuring consistency between them.

The form of the distribution of targets that we study is of the union of two independent components: $f^u_{t|t}(X)$ representing targets that so-far remain undetected[3] and $f^p_{t|t}(X)$ representing targets that have been previously detected and are thus under track. From these components we can reconstitute the full distribution as:

$$f_{t|t'}(X) = \sum_{Y \subseteq X} f^u_{t|t'}(Y) f^p_{t|t'}(X - Y) \tag{2}$$

[1] We use the term *resources* in the traditional operations research sense, *i.e.*, abstract quantities that are shared between tasks in an optimisation. In our case, resources correspond to measurements and tasks correspond to tracks, thus encoding the notion that each measurement may be used at most once in any association event.

[2] We leave off conditioning on the measurement set history $Z^t = (Z_1, \ldots, Z_t)$ throughout, as this is implicit in the second subscript $f_{t|t}$.

[3] *i.e.*, they have *never* previously been detected; this meaning will subsequently be altered in Section 6.

where $t' = t - 1$ for the one-step prediction, and $t' = t$ for the updated distribution. We digress momentarily to discuss the concept of maintaining a distribution of targets that have never been detected. While this may be troubling to some tracking practitioners, it has been suggested previously in [6], and studied extensively in search theory [12].[4] The reason for its inclusion is that when new targets are born, they are not necessarily detected in the first instance. Therefore, there is some probability that they will move (and potentially die) prior to being detected, which may occur several time steps later. If the initial intensity function of undetected targets, the birth rate of new targets, the probability of target death and the probability of detection are all spatially homogeneous within the region of interest, the intensity function of undetected targets will remain homogeneous through time and will converge to a constant value. However, realistically these quantities are non-homogeneous and time-varying for reasons including:

- Spatially non-homogeneous arrival rates of new targets, *e.g.*, at airports and at the boundary of the surveillance region (including due to sensor platform motion)
- Non-homogeneous target death probability (for similar reasons)
- Inherently non-homogenous, time-varying sensor detection performance (*e.g.*, performance variation with range, moving sensors, and agile sensors such as electronically scanned array radar)

Even in the case in which all these quantities are time-stationary, maintenance of a time-varying distribution of undetected targets can speed track initiation at system start-up, *i.e.*, as the system transitions from the steady state condition in which sensors are not functioning (where there are many undetected targets and no targets under track), to the steady state condition in which sensors are functioning (with few undetected targets and most targets under track). Assuming a non-unity probability of detection, many tracks will take several time steps to be detected during this transient; the undetected target distribution follows the expected distribution of newly detected targets over time to provide a posterior probability that each measurement is a newly detected target, incorporating knowledge of the history of sensor operation.

Returning to our derivation, the PPP representing undetected targets is:

$$f^u_{t|t}(X) \propto \prod_{x \in X} \lambda^u_{t|t}(x) \tag{3}$$

The component representing previously detected targets is:

$$f^p_{t|t}(\{x_1, \ldots, x_n\}) \propto \sum_{\alpha,a,b} \prod_{i \in \mathcal{T}_t} \left( w^{i,a^i}_{t|t} f^{i,a^i}_{t|t}(X_{\alpha(i)}) \prod_{j \in \mathcal{R}_t} \psi^{i,j}_r(a^i, b^j) \right) \tag{4}$$

[4]Our work in this area commenced with [13], which uses a PPP distribution of undetected targets to tractably solve non-myopic dwell time optimisation problems for new target search with an agile beam radar.

where

$$X_{\alpha(i)} \triangleq \begin{cases} \{x_{\alpha(i)}\}, & \alpha(i) > 0 \\ \emptyset, & \alpha(i) = 0 \end{cases}$$

The sum in (4) is over all functions $\alpha : \mathcal{T}_t \to \{0, \ldots, n\}$ such that $\{1, \ldots, n\} \subseteq \alpha(\mathcal{T}_t)$ and if $\alpha(i) > 0, i \neq j$ then $\alpha(i) \neq \alpha(j)$[5] (*i.e.*, there is exactly one track mapped to each $x_i,\ i \in \{1, \ldots, n\}$), and over the Cartesian products

$$a \in \prod_{i \in \mathcal{T}_t} \mathcal{H}_t^i, \quad b \in \prod_{j \in \mathcal{R}_t} \mathcal{T}_t$$

The resource tables $\psi_r^{i,j}(\cdot, \cdot)$ in (4) represent data association, *i.e.*, the constraint that two tracks cannot utilise the same measurement under any given event. The form may appear to be mysterious now; it will be explained further in Sections 2.2 and 3.

Finally, note that $f_{t|t}$ can be written in probability generating functional (PGFl) [1, p372] form as

$$G_{t|t}[h] = G_{t|t}^u[h] \cdot G_{t|t}^p[h]$$

where [1, p373]

$$G_{t|t}^u[h] \propto \exp\{\lambda_{t|t}^u[h]\}, \quad \lambda_{t|t}^u[h] = \int \lambda_{t|t}^u(x) h(x) \mathrm{d}x$$

and $G_{t|t}^p[h]$ is the PGFl of $f_{t|t}^p(X)$. Since, conditioned on a joint hypothesis $a = (a_1, \ldots, a_{n_t})$, this distribution is multi-target multi-Bernoulli, we can write this via linearity of the PGFl (see Appendix A) as: [1, p374]

$$G_{t|t}^p[h] = \sum_{a,b} \prod_{i \in \mathcal{T}_t} \left( w_{t|t}^{i,a^i} G_{t|t}^{i,a^i}[h] \prod_{j \in \mathcal{R}_t} \psi_r^{i,j}(a^i, b^j) \right) \tag{5}$$

where $G_{t|t}^{i,a}[h] = 1 - q_{t|t}^{i,a} + q_{t|t}^{i,a} \int h(x) f_{t|t}^{i,a}(x) \mathrm{d}x$ is the PGFl of $f_{t|t}^{i,a}(X)$.

## 2.1 Prediction step

The dynamics model described in the assumptions at the beginning of Section 2 corresponds to the case IV model in [1, p474]:

$$G_{t+1|t}[h|X'] \propto (1 - P^{\mathrm{s}} + P^{\mathrm{s}} p_h)^{X'} \cdot \exp\{\lambda^{\mathrm{b}}[h]\}$$

where, in accordance with notation from [1],

$$(1 - P^{\mathrm{s}} + P^{\mathrm{s}} p_h)^{X'} = \prod_{x' \in X'} [1 - P^{\mathrm{s}}(x') + P^{\mathrm{s}}(x') p_h(x')]$$

$$p_h(x') = \int f_{t+1|t}(x|x') h(x) \mathrm{d}x$$

[5]Note that if $n > n_t = |\mathcal{T}_t|$ then there is no such function, so the sum will be zero.

Then by [1, p529],

$$
\begin{aligned}
G_{t+1|t}[h] &\propto \exp\{\lambda^{\mathrm{b}}[h]\} \cdot G_{t|t}[1 - P^{\mathrm{s}} + P^{\mathrm{s}} p_h] \\
&= \exp\{\lambda^{\mathrm{b}}[h]\} \cdot G^u_{t|t}[1 - P^{\mathrm{s}} + P^{\mathrm{s}} p_h] \cdot G^p_{t|t}[1 - P^{\mathrm{s}} + P^{\mathrm{s}} p_h]
\end{aligned} \tag{6}
$$

The leading terms correspond directly to the prediction step in the PHD [14]:

$$
\begin{aligned}
G^u_{t+1|t}[h] &\propto \exp\{\lambda^{\mathrm{b}}[h]\} \cdot G^u_{t|t}[1 - P^{\mathrm{s}} + P^{\mathrm{s}} p_h] \\
&\propto \exp\{\lambda^{\mathrm{b}}[h] + \lambda^u_{t|t}[1 - P^{\mathrm{s}} + P^{\mathrm{s}} p_h]\} \\
&\propto \exp\{\lambda^{\mathrm{b}}[h] + \lambda^u_{t|t}[P^{\mathrm{s}} p_h]\} \\
&= \exp\{\lambda^u_{t+1|t}[h]\}
\end{aligned}
$$

where $\lambda^u_{t+1|t}[h] \triangleq \lambda^{\mathrm{b}}[h] + \lambda^u_{t|t}[P^{\mathrm{s}} p_h]$, so that

$$
\lambda^u_{t+1|t}(x) = \lambda^{\mathrm{b}}(x) + \int f_{t+1|t}(x|x') P^{\mathrm{s}}(x') \lambda^u_{t|t}(x') \mathrm{d}x' \tag{7}
$$

Consider now the term in (6) representing previously detected targets, $G^p_{t+1|t}[h] \triangleq G^p_{t|t}[1 - P^{\mathrm{s}} + P^{\mathrm{s}} p_h]$. Through the expression for the PGFl of $G^p_{t|t}[h]$ in (5) (which observes that, conditioned on a joint association hypothesis, the multi-target density is multi-Bernoulli), this can be calculated using the prediction step of the MeMBer:[6] [1, p675]

$$
\begin{aligned}
G^p_{t+1|t}[h] &\triangleq G^p_{t|t}[1 - P^{\mathrm{s}} + P^{\mathrm{s}} p_h] \\
&= \sum_{a,b} \prod_{i \in \mathcal{T}_t} \left( w^{i,a^i}_{t|t} G^{i,a^i}_{t|t}[1 - P^{\mathrm{s}} + P^{\mathrm{s}} p_h] \prod_{j \in \mathcal{R}_t} \psi^{i,j}_r(a^i, b^j) \right) \\
&= \sum_{a,b} \prod_{i \in \mathcal{T}_t} \left( w^{i,a^i}_{t|t} G^{i,a^i}_{t+1|t}[h] \prod_{j \in \mathcal{R}_t} \psi^{i,j}_r(a^i, b^j) \right)
\end{aligned} \tag{8}
$$

where

$$
\begin{aligned}
G^{i,a}_{t+1|t}[h] &\triangleq G^{i,a}_{t|t}[1 - P^{\mathrm{s}} + P^{\mathrm{s}} p_h] \\
&= 1 - q^{i,a}_{t|t} + q^{i,a}_{t|t} \int [1 - P^{\mathrm{s}}(x')] f^{i,a}_{t|t}(x') \mathrm{d}x' + \\
&\quad + q^{i,a}_{t|t} \iint h(x) f_{t+1|t}(x|x') \mathrm{d}x P^{\mathrm{s}}(x') f^{i,a}_{t|t}(x') \mathrm{d}x' \\
&= 1 - q^{i,a}_{t+1|t} + q^{i,a}_{t+1|t} \int h(x) f^{i,a}_{t+1|t}(x) \mathrm{d}x
\end{aligned}
$$

[6] With target death but without target birth, as this is handled through the separate Poisson component $G^u_{t+1|t}[h]$.

where the final equality is made by defining:

$$q^{i,a}_{t+1|t} \triangleq q^{i,a}_{t|t} \int P^{\mathrm{s}}(x') f^{i,a}_{t|t}(x') \mathrm{d}x'$$

$$f^{i,a}_{t+1|t}(x) \triangleq \frac{\int f_{t+1|t}(x|x') P^{\mathrm{s}}(x') f^{i,a}_{t|t}(x') \mathrm{d}x'}{\int P^{\mathrm{s}}(x') f^{i,a}_{t|t}(x') \mathrm{d}x'}$$

To complete the definitions, we set $w^{i,a}_{t+1|t} \triangleq w^{i,a}_{t|t}$, and $f^{i,a}_{t+1|t}(X)$ to be the multi-target distribution corresponding to the PGFl $G^{i,a}_{t+1|t}[h]$:

$$f^{i,a}_{t+1|t}(X) = \begin{cases} 1 - q^{i,a}_{t+1|t}, & X = \emptyset \\ q^{i,a}_{t+1|t} f^{i,a}_{t+1|t}(x), & X = \{x\} \\ 0, & |X| \geq 2 \end{cases}$$

Consequently, the prediction step on the distribution of previously detected targets is implemented simply by processing each track separately with the above expressions.

## 2.2 Measurement update step

The measurement model described in the assumptions at the beginning of Section 2 corresponds to the case V model in [1, p422]:

$$G_{t+1}[g|X] \propto \exp\{\lambda^{\mathrm{fa}}[g]\}(1 - P^{\mathrm{d}} + P^{\mathrm{d}} p_g)^X$$

where

$$(1 - P^{\mathrm{d}} + P^{\mathrm{d}} p_g)^X \triangleq \prod_{x \in X} [1 - P^{\mathrm{d}}(x) + P^{\mathrm{d}}(x) p_g(x)]$$

$$p_g(x) \triangleq \int f(z|x) g(x) \mathrm{d}x$$

Subsequently, the joint PGFl of measurements and targets can be written as: [1, p531]

$$\begin{aligned} F[g,h] &\propto \exp\{\lambda^{\mathrm{fa}}[g]\} \cdot G_{t+1|t}[h(1 - P^{\mathrm{d}} + P^{\mathrm{d}} p_g)] \\ &= \exp\{\lambda^{\mathrm{fa}}[g]\} \cdot G^{u}_{t+1|t}[h(1 - P^{\mathrm{d}} + P^{\mathrm{d}} p_g)] \cdot G^{p}_{t+1|t}[h(1 - P^{\mathrm{d}} + P^{\mathrm{d}} p_g)] \\ &\propto \exp\{\lambda^{\mathrm{fa}}[g] + \lambda^{u}_{t+1|t}[h(1 - P^{\mathrm{d}} + P^{\mathrm{d}} p_g)]\} \cdot G^{p}_{t+1|t}[h(1 - P^{\mathrm{d}} + P^{\mathrm{d}} p_g)] \\ &= \exp\{\lambda^{u}_{t+1|t}[h(1 - P^{\mathrm{d}})]\} \cdot \exp\{\lambda^{\mathrm{fa}}[g] + \lambda^{u}_{t+1|t}[h P^{\mathrm{d}} p_g]\} \cdot \\ &\quad \cdot G^{p}_{t+1|t}[h(1 - P^{\mathrm{d}} + P^{\mathrm{d}} p_g)] \end{aligned} \tag{9}$$

By [1, p530],

$$G_{t+1|t+1}[h] \propto \left. \frac{\delta F}{\delta Z_{t+1}}[g,h] \right|_{g=0}$$

Since the first term in (9) does not involve measurements (*i.e.*, the measurement functional $g$ does not appear), we can separate this into terms

$$G_{t+1|t+1}[h] = G^u_{t+1|t+1}[h] \cdot G^p_{t+1|t+1}[h]$$

where the distribution of undetected targets is equivalent to a PHD update in the absence of measurements: [14]

$$\begin{aligned} G^u_{t+1|t+1}[h] &\propto \exp\{\lambda^u_{t+1|t}[h(1-P^{\mathrm{d}})]\} \\ &= \exp\{\lambda^u_{t+1|t+1}[h]\} \\ \lambda^u_{t+1|t+1}(x) &\triangleq [1-P^{\mathrm{d}}(x)]\lambda^u_{t+1|t}(x) \end{aligned} \tag{10}$$

We now turn to the distribution of detected targets (both those that have been detected previously, and those that are detected for the first time at $(t+1)$):

$$G^p_{t+1|t+1}[h] \propto \frac{\delta}{\delta Z}\left(\exp\{\lambda^{\mathrm{fa}}[g] + \lambda^u_{t+1|t}[hP^{\mathrm{d}}p_g]\} \cdot G^p_{t+1|t}[h(1-P^{\mathrm{d}}+P^{\mathrm{d}}p_g)]\right)\Big|_{g=0}$$

This can be evaluated via the chain rule: [1, p389]

$$\frac{\delta}{\delta Y}(F_1[h]\cdots F_n[h]) = \sum_{W_1\uplus\cdots\uplus W_n = Y} \frac{\delta F_1}{\delta W_1}[h]\cdots\frac{\delta F_n}{\delta W_n}[h] \tag{11}$$

where the notation $\uplus$ denotes that the sum is over all disjoint sets $W_1, \ldots, W_n$ such that $W_1 \cup \cdots \cup W_n = Y$. To begin, we consider the components required to evaluate (11). It can be easily shown that

$$\frac{\delta}{\delta Z}\exp\{\lambda[g]\}\Big|_{g=0} = \prod_{z\in Z}\lambda(z)$$

Consequently,

$$\begin{aligned} &\frac{\delta}{\delta Z}\exp\{\lambda^{\mathrm{fa}}[g] + \lambda^u_{t+1|t}[hP^{\mathrm{d}}p_g]\}\Big|_{g=0} \\ &\quad= \prod_{z\in Z}\left(\lambda^{\mathrm{fa}}(z) + \lambda^u_{t+1|t}[hP^{\mathrm{d}}f(z|\cdot)]\right) \\ &\quad= \prod_{z\in Z} w^z_{t+1|t+1}\left[(1-q^z_{t+1|t+1}) + q^z_{t+1|t+1}\int f^z_{t+1|t+1}(x)h(x)\mathrm{d}x\right] \\ &\quad= \prod_{z\in Z} w^z_{t+1|t+1} G^z_{t+1|t+1}[h] \end{aligned} \tag{12}$$

where

$$
\begin{aligned}
\lambda^u_{t+1|t}[hP^{\mathrm{d}}f(z|\cdot)] &\triangleq \int h(x)f(z|x)P^{\mathrm{d}}(x)\lambda^u_{t+1|t}(x)\mathrm{d}x \\
w^z_{t+1|t+1} &\triangleq \lambda^{\mathrm{fa}}(z) + \lambda^u_{t+1|t}[f(z|\cdot)P^{\mathrm{d}}] \\
q^z_{t+1|t+1} &\triangleq \frac{\lambda^u_{t+1|t}[f(z|\cdot)P^{\mathrm{d}}]}{\lambda^{\mathrm{fa}}(z) + \lambda^u_{t+1|t}[f(z|\cdot)P^{\mathrm{d}}]} \\
f^z_{t+1|t+1}(x) &\triangleq \frac{f(z|x)P^{\mathrm{d}}(x)\lambda^u_{t+1|t}(x)}{\lambda^u_{t+1|t}[f(z|\cdot)P^{\mathrm{d}}]} \\
G^z_{t+1|t+1}[h] &\triangleq (1 - q^z_{t+1|t+1}) + q^z_{t+1|t+1}f^z_{t+1|t+1}[h]
\end{aligned} \tag{13}
$$

As shown in (8), $G^p_{t+1|t}[h]$ can be decomposed into a sum of multi-Bernoulli distributions, hence $G^p_{t+1|t+1}[h]$ will be composed of many Bernoulli updates of the form

$$
\begin{aligned}
w^{i,a,Z}_{t+1|t+1}G^{i,a,Z}_{t+1|t+1}[h] &= \frac{\delta}{\delta Z} w^{i,a}_{t+1|t}G^{i,a}_{t+1|t}[h(1-P^{\mathrm{d}}+P^{\mathrm{d}}p_g)]\bigg|_{g=0} \\
&= w^{i,a}_{t+1|t}\begin{cases} 1 - q^{i,a}_{t+1|t} + q^{i,a}_{t+1|t}f^{i,a}_{t+1|t}[h(1-P^{\mathrm{d}})], & Z = \emptyset \\ q^{i,a}_{t+1|t}f^{i,a}_{t+1|t}[hP^{\mathrm{d}}f(z|\cdot)], & Z = \{z\} \\ 0, & |Z| \geq 2 \end{cases}
\end{aligned} \tag{14}
$$

so that, in the case where $Z = \emptyset$, we find

$$
\begin{aligned}
G^{i,a,\emptyset}_{t+1|t+1}[h] &= 1 - q^{i,a,\emptyset}_{t+1|t+1} + q^{i,a,\emptyset}_{t+1|t+1}f^{i,a,\emptyset}_{t+1|t+1}[h] \\
w^{i,a,\emptyset}_{t+1|t+1} &\triangleq w^{i,a}_{t+1|t}(1 - q^{i,a}_{t+1|t} + q^{i,a}_{t+1|t}f^{i,a}_{t+1|t}[1-P^{\mathrm{d}}]) \\
q^{i,a,\emptyset}_{t+1|t+1} &\triangleq \frac{q^{i,a}_{t+1|t}f^{i,a}_{t+1|t}[1-P^{\mathrm{d}}]}{1 - q^{i,a}_{t+1|t} + q^{i,a}_{t+1|t}f^{i,a}_{t+1|t}[1-P^{\mathrm{d}}]} \\
f^{i,a,\emptyset}_{t+1|t+1}(x) &\triangleq \frac{[1-P^{\mathrm{d}}(x)]f^{i,a}_{t+1|t}(x)}{f^{i,a}_{t+1|t}[1-P^{\mathrm{d}}]}
\end{aligned} \tag{15}
$$

and when $Z = \{z\}$, we find

$$
\begin{aligned}
G^{i,a,\{z\}}_{t+1|t+1}[h] &= 1 - q^{i,a,\{z\}}_{t+1|t+1} + q^{i,a,\{z\}}_{t+1|t+1}f^{i,a,\{z\}}_{t+1|t+1}[h] \\
w^{i,a,\{z\}}_{t+1|t+1} &\triangleq w^{i,a}_{t+1|t}q^{i,a}_{t+1|t}f^{i,a}_{t+1|t}[f(z|\cdot)P^{\mathrm{d}}] \\
q^{i,a,\{z\}}_{t+1|t+1} &\triangleq 1 \\
f^{i,a,\{z\}}_{t+1|t+1}(x) &\triangleq \frac{f(z|x)P^{\mathrm{d}}(x)f^{i,a}_{t+1|t}(x)}{f^{i,a}_{t+1|t}[f(z|\cdot)P^{\mathrm{d}}]}
\end{aligned} \tag{16}
$$

In order to apply the chain rule (11), we seek a convenient form for a summation over all subsets of $Z_{t+1}$. Lemma 1 shows that this can be achieved using a form similar to that in (4). The motivation for using this form will be discussed in Section 3.

**Lemma 1.** *Suppose we are given set functions $F_i(Z)$, $i \in \{0, ..., n\}$, such that $F_0(Z) = \prod_{z\in Z} f_0(z)$ and if $i \geq 1$ and $|Z| \geq 2$ then $F_i(Z) = 0$. If $Z = \{z^1, \ldots, z^m\}$, then*

$$\sum_{W_0 \uplus \cdots \uplus W_n = Z} F_0(W_0) \cdots F_n(W_n)$$

$$= \sum_{a,b} \left( \prod_{j|b^j=0} f_0(z^j) \right) \cdot \prod_{i=1}^{n} \left( F_i(Z^{a^i}) \prod_{j=1}^{m} \tilde{\psi}_r^{i,j}(a^i, b^j) \right)$$

*where the sum over $(a, b)$ is over all $a = (a^1, \ldots, a^n)$ such that $a^i \in \{0, \ldots, m\}\ \forall\, i$ and all $b = (b^1, \ldots, b^m)$ such that $b^j \in \{0, \ldots, n\}\ \forall\, j$, and*

$$Z^{a^i} \triangleq \begin{cases} \emptyset, & a^i = 0 \\ z^{a^i}, & a^i > 0 \end{cases}$$

$$\tilde{\psi}_r^{i,j}(a^i, b^j) \triangleq \begin{cases} 0, & b^j = i, a^i \neq j \text{ or } a^i = j, b^j \neq i \\ 1, & \text{otherwise} \end{cases}$$

*Proof.* We show that there is a one-to-one equivalence between non-zero terms in the LHS and RHS expressions. First, consider a non-zero LHS term, *i.e.*, a choice of $(W_0, \ldots, W_n)$ (with $W_i$ mutually disjoint and $W_0 \cup \cdots \cup W_n = Z$) for which $F_0(W_0) \cdots F_n(W_n) \neq 0$. For each $i \in \{1, \ldots, n\}$ we either have $W_i = \{z^{a^i}\}$ for some $a^i \in \{1, \ldots, m\}$, or $W_i = \emptyset$ or in which case let $a^i = 0$. Then $W_i = Z^{a^i}\ \forall\ i \in \{1, \ldots, n\}$. For each $j \in \{1, \ldots, m\}$ let $\mathcal{B}^j = \{i | a^i = j\}$. If $\mathcal{B}^j = \emptyset$ then set $b^j = 0$. If $\mathcal{B}^j = \{i\}$ then set $b^j = i$. Note that $|\mathcal{B}^j| \leq 1$ as $W_i = Z^{a^i}$ and the sets $W_i$ are disjoint. Then since $W_0 = Z \backslash (W_1 \cup \cdots \cup W_n)$, we must have $W_0 = \{z^j | b^j = 0\}$. By construction, $\tilde{\psi}_r^{i,j}(a^i, b^j) = 1\ \forall\ (i, j)$, hence there is a RHS term that is equal to the LHS term under consideration. It is clear from the construction that any change to $(W_0, \ldots, W_n)$ will produce a different $(a, b)$, so the mapping is injective.

Now consider a non-zero RHS term, *i.e.*, a choice of $a = (a^1, \ldots, a^n)$ and $b = (b^1, \ldots, b^m)$ such that $\tilde{\psi}_r^{i,j}(a^i, b^j) = 1\ \forall\ (i, j)$. Let $W_0 = \{z^j | b^j = 0\}$, and $W_i = Z^{a^i}$ for $i \in \{1, \ldots, n\}$. Seeking a contradiction, we suppose that the sets $(W_0, \ldots, W_n)$ are not disjoint. If $W_0 \cap W_i \neq \emptyset$ then we have $\tilde{\psi}_r^{i,j}(a^i, b^j) = 0$ for $j = a^i$ (since $b^j = 0$) which is a contradiction. Similarly, if $W_{i'} \cap W_i \neq \emptyset$, $i' \neq i$, then for $j = a^i$, we must have either $\tilde{\psi}_r^{i,j}(a^i, b^j) = 0$ or $\tilde{\psi}_r^{i',j}(a^{i'}, b^j) = 0$, which is a contradiction. Thus $(W_0, \ldots, W_n)$ are disjoint, and the LHS term involving $(W_0, \ldots, W_n)$ is equal to the RHS term under consideration.

Finally, suppose that $(a, b)$ and $(\tilde{a}, \tilde{b})$ both map to $(W_0, \ldots, W_n)$. Then we must have $a = \tilde{a}$ and $\{j | b^j = 0\} = \{j | \tilde{b}^j = 0\}$. Suppose $b^j = i \neq \tilde{b}^j$. Then since $a^i = \tilde{a}^i$, either $\tilde{\psi}_r^{i,j}(a^i, b^j) = 0$ or $\tilde{\psi}_r^{i,j}(\tilde{a}^i, \tilde{b}^j) = 0$, hence one of these two terms is zero. □

The following corollary is a direct consequence of the chain rule of (11), followed by an application of Lemma 1.

**Corollary 1.** *If the a priori distribution of pre-existing targets is multi-Bernoulli, i.e.,*

$$G^p_{t+1|t}[h] = \prod_{i \in \mathcal{T}_t} w^i_{t+1|t} G^i_{t+1|t}[h]$$

*where $G^i_{t+1|t}[h]$ ($i \in \mathcal{T}_t$) is a Bernoulli distribution (initially assume $w^i_{t+1|t} = 1$, but we retain the weights to permit different values), then the distribution updated with $Z_{t+1} = \{z^1, \ldots, z^{m_{t+1}}\}$ is*

$$G^p_{t+1|t+1}[h] \propto \sum_{a,b} \left( \prod_{j|b^j=0} w^{z^j}_{t+1|t+1} G^{z^j}_{t+1|t+1}[h] \right) \cdot \\ \cdot \prod_{i=1}^{n_t} \left( w^{i,Z^{a^i}}_{t+1|t+1} G^{i,Z^{a^i}}_{t+1|t+1}[h] \prod_{j=1}^{m_{t+1}} \tilde{\psi}_r^{i,j}(a^i, b^j) \right)$$

Since the predicted distribution in (8) is a mixture (*i.e.*, linear combination) of multi-Bernoulli distributions, the desired update follows directly, as given in the following corollary.

**Corollary 2.** *Assume $G^p_{t+1|t}[h]$ is as given in (8). Then*

$$G^p_{t+1|t+1}[h] \propto \sum_{\tilde{a},\tilde{b}} \prod_{i \in \mathcal{T}_{t+1}} \left( w^{i,\tilde{a}^i}_{t+1|t+1} G^{i,\tilde{a}^i}_{t+1|t+1}[h] \prod_{j \in \mathcal{R}_{t+1}} \psi_r^{i,j}(\tilde{a}^i, \tilde{b}^j) \right)$$

*where $\mathcal{T}_{t+1} = \{1, \ldots, n_{t+1}\}$, $n_{t+1} = n_t + m_{t+1}$, $\mathcal{R}_{t+1} = \mathcal{R}_t \cup \{(t+1, j) | j \in \{1, \ldots, m_{t+1}\}\}$, the sum over $\tilde{a}$ and $\tilde{b}$ is over all*

$$\tilde{a} \in \prod_{i \in \mathcal{T}_{t+1}} \mathcal{H}^i_{t+1}, \quad \tilde{b} \in \prod_{j \in \mathcal{R}_{t+1}} \mathcal{T}_{t+1}$$

*For existing tracks $i \in \{1, \ldots, n_t\}$, $\mathcal{H}^i_{t+1} = \mathcal{H}^i_t \cup \{(a^i \cup \{(t+1, j)\}) | a^i \in \mathcal{H}^i_t,\ j \in \{1, \ldots, m_{t+1}\}\}$; for each $\tilde{a}^i = a^i \in \mathcal{H}^i_t$, $w^{i,a^i}_{t+1|t+1} = w^{i,a^i,\emptyset}_{t+1|t+1}$ and $G^{i,a^i}_{t+1|t+1}[h] = G^{i,a^i,\emptyset}_{t+1|t+1}[h]$ from (15); and for each $\tilde{a}^i = (a^i \cup \{(t+1, j)\}) \in \mathcal{H}^i_{t+1}$, $w^{i,\tilde{a}^i}_{t+1|t+1} = w^{i,a^i,\{z^j\}}_{t+1|t+1}$ and $G^{i,\tilde{a}^i}_{t+1|t+1}[h] = G^{i,a^i,\{z^j\}}_{t+1|t+1}[h]$ from (16).*

*For new tracks $i = n_t + j$, $j \in \{1, \ldots, m_{t+1}\}$, $\mathcal{H}^i_{t+1} = \{(\emptyset, \{(t+1, j)\})\}$, $w^{i,\emptyset}_{t+1|t+1} = 1$, $G^{i,\emptyset}_{t+1|t+1}[h] = 1$, $w^{i,\{(t+1,j)\}}_{t+1|t+1} = w^{z^j}_{t+1|t+1}$ and $G^{i,\{(t+1,j)\}}_{t+1|t+1}[h] = G^{z^j}_{t+1|t+1}[h]$ from (13).*

## 2.3 Initialisation

Initialisation can take advantage of whatever prior information is available. The natural choice is to set $\lambda^u_{0|0}(x)$ to be the steady state intensity of undetected targets in the absence of sensor measurements, *i.e.*, the $\lambda^u(x)$ which satisfies

$$\lambda^u(x) = \lambda^b(x) + \int f_{t+1|t}(x|x')P^{\mathrm{s}}(x')\lambda^u(x')\mathrm{d}x'$$

This is the steady state distribution of targets induced by target birth (with intensity $\lambda^{\mathrm{b}}(x)$), movement (with the transition PDF $f_{t+1|t}(x|x')$) and death (via the survival probability $P^{\mathrm{s}}(x)$). If these quantities are modelled sufficiently accurately, this distribution should provide a faithful representation of the expected number of targets present in the scene. Since there is no detection history and hence no tracks, we set $\mathcal{T}_0 = \emptyset$ and $\mathcal{R}_0 = \emptyset$.

## 2.4 Representation of undetected target distribution

So far we have said little about the distribution of undetected targets, as defined by (7) and (10). There is little prior work in which this has not been assumed to be homogeneous. We propose discretising the state space in order to represent this distribution via a grid filter. Grid-based representations (*e.g.*, [15]) have fallen out of vogue for most estimation applications as they are generally inefficient. In particular, in any problem where the probability distributions are peaked, grid representations result in large computing resources being spent on the vast majority of cells with near-zero probability density. In comparison, the preferred sample-based methods [16] allow computing resources to be focussed on the region with significant probability mass. However, in the present application, the Poisson intensity function of undetected targets is, by nature, both diffuse and smooth. Thus a grid-based representation is efficient (since most grid points will have comparable intensity), and, furthermore, a coarse discretisation suffices, such that the grid-based method is tractable. Conversely, sample-based methods using any reasonable number of samples would be likely to exhibit poor sample coverage in the vicinity of some cells.

## 2.5 Comments and relationship to existing work

At this point, it may seem that little has been achieved other than an alternative derivation of the full Bayes RFS filter. The derivation closely parallels [2], with the addition of a PPP component of undetected targets, which was suggested previously in [6]. The major purpose of the derivation was to incorporate the somewhat obscure form of (4). In Section 3, we show how this form permits application of the method described in [4,5], which is the major contribution of this paper. Graphical models, upon which our approximation is built, were first applied to tracking in [17–19], which focused on distributed sensor networks.

Because of the hybrid (PPP, MeMBer) representation, there are interesting relationships between the proposed algorithm and both the PHD and MeMBer filters. The prediction equation for the PPP component (7) is exactly the PHD prediction equation, and the update equation for the PPP component (10) is the PHD update for the case in which there are no measurements. In Section 6 we propose a concept of *recycling*, which reduces the number of tracks that need to be maintained by representing low probability of existence tracks via the PPP component. In [14], Mahler interprets the PPP as being a projection of the updated multi-target distribution onto the subspace of PPP distributions. Not surprisingly then, if we project *all* tracks onto the PPP representation, our filter reduces to the PHD. The concept of projecting a *subset* of tracks onto the PPP representation permits us to focus computational resources on tracks with a non-negligible probability of existence while still being able to gradually accrue confidence in low SNR tracks (*e.g.*, in low probability of detection conditions). We expect that this structure will be most useful in track before detect (TkBD) problems; extending our derivation to this case is a topic of future work.

The relationship between the previously detected track portion of the proposed algorithm and the MeMBer filter is also interesting; this relationship is very similar to that between M-C and MeMBer, and is explored in Fig. 1. The figure shows the hypothesis tree structure for the tracks maintained by the proposed method and the MeMBer. In the proposed method, at each time each existing leaf hypothesis is updated with each new measurement to form new leaves. Additionally, new tracks are formed representing the hypothesis that the measurement is a newly detected target. Consistency constraints are enforced between trees in the calculation of approximate marginal probabilities via the method described in Section 3. In the MeMBer, each existing track is retained, updating each leaf assuming a missed detection. An additional track is formed for each new measurement, incorporating hypotheses that the measurement is a new target, or an extension of each previous track. Each track is treated as being independent, *i.e.*, no consistency constraints are enforced.

The probabilistic structure implied by the hypothesis trees is such that one may choose one leaf from each tree[7] in any association event. The MeMBer places priority on the constraints that each measurement in the latest scan can correspond to at most one target, and re-forms the hypothesis trees such that these constraints are implicit in the tree structure (*i.e.*, all hypotheses updated with the same measurement in the latest scan are in the same track). Consequently, continuity of association history is lost; *e.g.*, association events in which the same historical track is updated with two different measurements are legal, since these reside in different tracks. Conversely, in the proposed method (and M-C), continuity of tracks is maintained, but there are leaves of different trees corresponding to the different tracks being updated with the same measurement. In problems involving closely spaced targets, these constraints cannot be ignored, which brings about the complexity of the proposed method, M-C and MHT, as well as the problem of coalescence. Our solution to the problem of computational complexity is detailed in Section 3, and our solution to the problem of coalescence is described in Section 4.

[7] *i.e.*, one association history hypothesis for each track

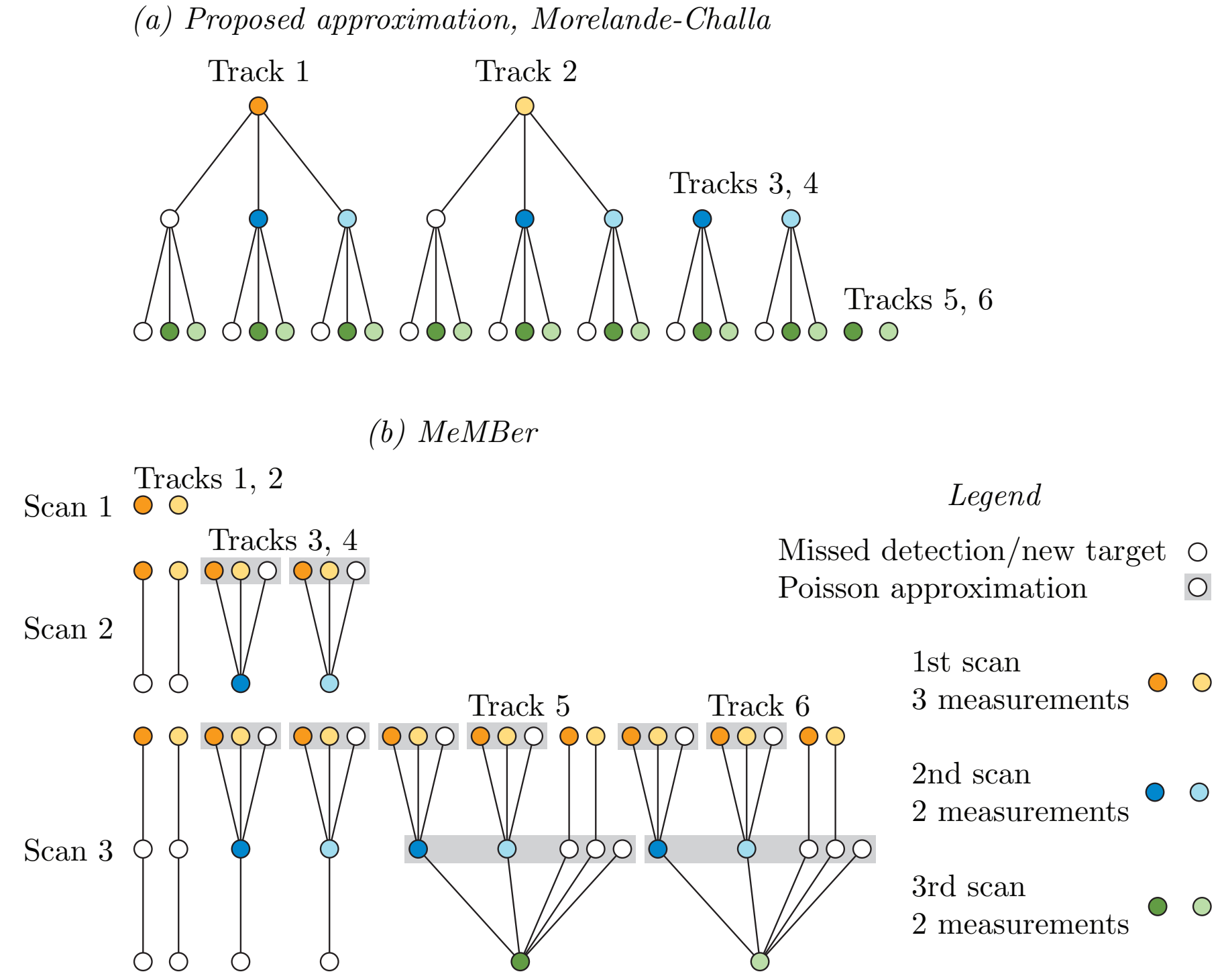

Figure 1: Hypothesis trees for (a) proposed method (same as that of M-C), and (b) MeMBer.

One significant advantage of the structure proposed is that, in the case of well-separated targets, it reduces to a series of parallel JoTT filters, which is optimal in these conditions. In contrast, the MeMBer filter still splits the tracks into hypotheses that are not updated, and tracks that are updated with each measurement. As described in the derivation of the MeMBer [1, p668, pp678–681], the impact of this approximation will be low if the probability of detection is high and the false alarm probability is low (since the legacy track will have low weight), but it will be significant in medium to low probability of detection conditions.

Two variations of the JoTT filter to address cases with association conflicts were described in [20]. The first uses a linear multi-target approximation (*e.g.*, [21]), which, for each track, approximates all other tracks as being represented as a Poisson intensity. The other is effectively a global nearest neighbours filter, modified to incorporate target existence. Our method is closely related to these, and performance will be compared in future experiments.

Finally, we note that it is possible to use our approximate method to enforce constraints from *previous* measurement scans, rather than the most recent scan. This leads to an alternative approximation that uses the hypothesis structure from the MeMBer, but uses the graphical model-based method to approximate the weights within each track. This method is described in Section 5. We also note that the original MeMBer filter in [1] has been shown to exhibit cardinality bias, which was subsequently addressed in [22].

# 3 Belief propagation marginal track filter (BPMTF)

Consider the form of (4), and observe that it can be written equivalently as:

$$f^p_{t|t}(X) = \sum_{\alpha} \sum_{a} p_{t|t}(a) \prod_{i \in \mathcal{T}} f^{i,a^i}_{t|t}(X_{\alpha(i)}) \tag{17}$$

An obvious simplification is to approximate the joint association event distribution $p_{t|t}(a)$ by the product of its marginal distributions $\prod_{i \in \mathcal{T}_t} p^i_{t|t}(a^i)$. This leads directly leads directly to an approximation of the multi-target distribution:

$$\tilde{G}^p_{t|t}[h] = \prod_{i \in \mathcal{T}_t} G^i_{t|t}[h] \tag{18}$$

or equivalently

$$\tilde{f}^p_{t|t}(X) = \sum_{\alpha} \prod_{i \in \mathcal{T}_t} f^i_{t|t}(X_{\alpha(i)}) \tag{19}$$

where

$$G^i_{t|t}[h] = \sum_{a^i \in \mathcal{H}^i_{t|t}} p^i_{t|t}(a^i) G^{i,a^i}_{t|t}[h] \tag{20}$$

$$f^i_{t|t}(X_{\alpha(i)}) = \sum_{a^i \in \mathcal{H}^i_{t|t}} p^i_{t|t}(a^i) f^{i,a^i}_{t|t}(X_{\alpha(i)}) \tag{21}$$

This is precisely the multi-Bernoulli form of [1, p656], and the use of marginal association probabilities is equivalent to that of [2]. The difficulty in implementing this approximation is in the calculation of the marginal association probabilities $p^i_{t|t}(a^i)$; in general this calculation requires summation over all joint association events.

This section discusses the framework of graphical models; in Section 3.4 we show how we can employ emerging approximate methods from within this framework to approximate these marginal probabilities. A significant amount of background material is included to provide a level of confidence that the method is based on a principled, convex optimisation-based approximation to the exact quantities.

We refer to the filter of the above form, replacing the exact marginal distributions with their approximations calculated using belief propagation, as the belief propagation marginal track filter (BPMTF).

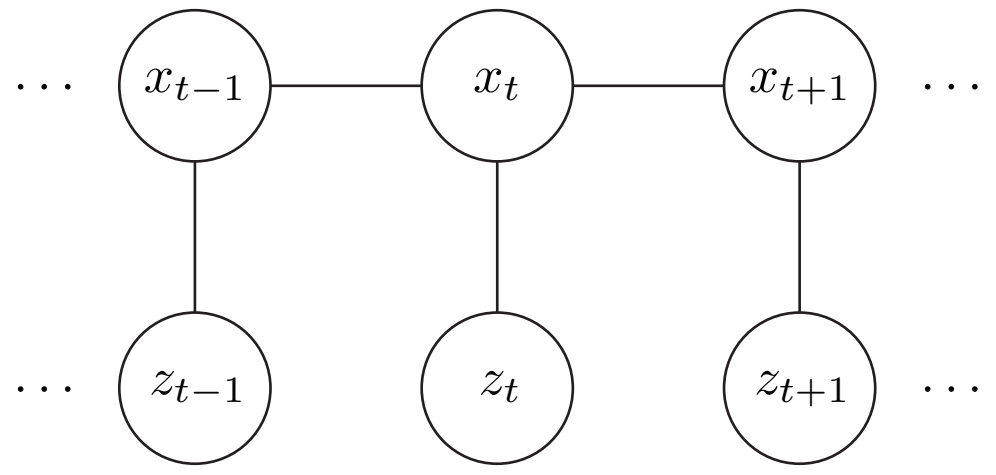

Figure 2: Section of a graphical model depiction of an HMM. The hidden state at time $t$ is $x_t$, and the observation at time $t$ is $z_t$. One choice of factors would be $\psi_{x_0}(x_0) = p(x_0)$, $\psi_{(x_{t-1},x_t)}(x_{t-1}, x_t) = p(x_t|x_{t-1})$ and $\psi_{(x_t,z_t)}(x_t, z_t) = p(z_t|x_t)$.

### 3.1 Graphical models and belief propagation

Belief propagation (BP) [23–25] is a generalisation of common inference algorithms on Markov chains such as forwards-backwards [25–27] in discrete hidden Markov models (HMMs) and the Kalman filter[8] [29,30] in Gauss-Markov chains. Systems of random variables are represented as an undirected graph $(\mathcal{G}, \mathcal{E})$, where the vertices of the graph $v \in \mathcal{V}$ represent random variables, and the edges $e \in \mathcal{E} \subseteq \mathcal{V} \times \mathcal{V}$ represent probabilistic dependencies between random variables.[9] For example, the graphical depiction of an HMM is shown in Fig. 2.

For clarity, we restrict ourselves to pairwise graphical models, although this is not required.[10] The joint probability distribution of the variables $x = (x_v)_{v \in \mathcal{V}}$ can be written as:

$$p(x) \propto \prod_{v \in \mathcal{V}} \psi_v(x_v) \prod_{(v_1,v_2) \in \mathcal{E}} \psi_{(v_1,v_2)}(x_{v_1}, x_{v_2}) \tag{22}$$

where $\psi_v(\cdot)$ and $\psi_{(v_1,v_2)}(\cdot, \cdot)$ are the factors that collectively make up the distribution. A graph encodes the conditionally independence structure in a network of random variables; *e.g.*, if all paths from vertices $\mathcal{A} \subset \mathcal{V}$ to $\mathcal{B} \subset \mathcal{V}$ pass through the vertex set $\mathcal{C} \subset \mathcal{V}$, then $\mathcal{A}$ and $\mathcal{B}$ are conditionally independent given $\mathcal{C}$. In the case of Fig. 2, this simply corresponds to the Markovianity property, and the property that the observation at a given time is conditionally independent of all other states and observations given the state at the same time.

If a graph is acyclic,[11] then inference may be performed exactly by iterating

[8] More precisely, the fixed interval Kalman smoother [28].

[9] Since we are dealing with undirected graphs, if $(v_1, v_2) \in \mathcal{E}$, then $(v_2, v_1) \in \mathcal{E}$.

[10] As described in [31, p289], any graphical model can be converted to a pairwise model via a simple transformation; factor graphs [32] provide a systematic mechanism which uses that transformation. Alternatively, the distribution may be written as the product of factors on maximal cliques.

[11] *i.e.*, there is at most one path between any two vertices.

the following equation for all $(v_1, v_2) \in \mathcal{E}$:[12] [23–25,31]

$$\mu_{v_1 \to v_2}(x_{v_2}) = \int \left[ \prod_{v \neq v_2 | (v, v_1) \in \mathcal{E}} \mu_{v \to v_1}(x_{v_1}) \right] \psi_{v_1}(x_{v_1}) \psi_{(v_1,v_2)}(x_{v_1}, x_{v_2}) \mathrm{d}x_{v_1} \tag{23}$$

The quantities $\mu_{v_1 \to v_2}(\cdot)$ are referred to as *messages*, since they are exchanged between neighbouring vertices. Upon convergence (*i.e.*, when subsequent iterates of all messages are identical), the marginal distribution of any vertex $v \in \mathcal{V}$ may be recovered as:

$$p(x_v) \propto \left[ \prod_{v' | (v, v') \in \mathcal{E}} \mu_{v' \to v}(x_v) \right] \psi_v(x_v) \tag{24}$$

and the pairwise distribution of any edge $(v_1, v_2) \in \mathcal{E}$ may be recovered as:

$$p(x_{v_1}, x_{v_2}) \propto \psi_{v_1}(x_{v_1}) \psi_{v_2}(x_{v_2}) \psi_{(v_1,v_2)}(x_{v_1}, x_{v_2}) \cdot \\ \cdot \left[ \prod_{v' \neq v_2 | (v_1, v') \in \mathcal{E}} \mu_{v' \to v_1}(x_{v_1}) \right] \left[ \prod_{v' \neq v_1 | (v_2, v') \in \mathcal{E}} \mu_{v' \to v_2}(x_{v_2}) \right] \tag{25}$$

For obvious reasons, this is also referred to as the *sum-product* algorithm; as previously stated, it may be seen to be a generalisation of the fixed interval Kalman smoother [28], and of the forwards-backwards algorithm for inference in HMMs [25–27]. The related *max-product* algorithm replaces the integral in (23) with a maximisation operation, providing a generalisation of the Viterbi algorithm [25,27,33] to trees just as sum-product generalises the forwards-backwards algorithm. Algebraically, this simply replaces the $(+, \cdot)$ semi-ring with the $(\max, \cdot)$ or, working in negative log-space, the $(\min, +)$ semi-ring.

Inference on cyclic graphs can be conducted optimally using a similar algorithm on a modified graph referred to as a *junction tree* [24,25]. Loosely speaking, the concept is to merge vertices together into hyper-vertices, until one obtains a graph which is a tree. In the discrete case, the complexity of this algorithm increases exponentially with the maximum hyper-vertex size.

## 3.2 Loopy belief propagation

While the algorithm above was derived for and is optimal for acyclic graphs, it can also be applied directly to cyclic graphs; this is referred to as loopy belief propagation (LBP). In the case of cyclic graphs, LBP is neither guaranteed to converge to the optimal result, nor to converge at all. Nevertheless, and perhaps surprisingly, it has exhibited excellent empirical performance in may practical problems [34]. For example, the popular iterative turbo decoding algorithm has been shown to be an instance of LBP [35].

[12] In discrete (or hybrid) cases, an appropriate counting (hybrid) measure is used.

The common wisdom is that the performance of LBP will be acceptable when there are few cycles, and when the coupling strength (*i.e.*, the degree of correlation/dependency between variables) in the cycles is weak [34]. Conversely, when there are many cycles involving strong coupling, performance is expected to be problematic. In recent years, however, there have emerged some dense, loopy structures for which performance is surprisingly good. The remarkable result that max-product LBP applied to a complete bipartite graph optimally solves the two dimensional assignment problem in time comparable to that of the auction algorithm was proven in [36]. Since then, authors from machine learning, statistical physics, information theory and tracking [4,5,37–40] have studied the properties of sum-product LBP as an approximation to the related #P-complete problem of calculating the matrix permanent. A by-product of this is algorithm is an approximation of the marginal association probabilities that are required for JPDA and (21). The accuracy of the approximate marginal association probabilities was studied in [4], and convergence was proven in [5]. A more general convergence proof was obtained via a different route in the parallel work [40,41]. The latter work also showed that the Bethe free energy is a convex lower bound to the Gibbs free energy for the problem of interest. Section 3.3 develops the theory required to understand this statement; Section 3.4 presents the model to which these results apply, and the results themselves are discussed in Section 3.4.3.

## 3.3 Gibbs free energy, Bethe free energy

The problem of finding the marginal distributions of vertices (random variables) in a graphical model can be posed as a *variational problem*, *i.e.*, as an optimisation. The common optimisation is one involving the Kullback-Leibler (KL) divergence [42,43]:

$$\underset{q}{\text{minimise}} \int q(x) \log \frac{q(x)}{p(x)} \mathrm{d}x \tag{26}$$

It is well-known that (*e.g.*, [42]) in the above problem, the unique minimum value of zero is attained by setting[13] $q(x) = p(x)$. Manipulating the objective, we obtain:

$$\begin{aligned} J(q) &\triangleq \int q(x) \log \frac{q(x)}{p(x)} \mathrm{d}x \\ &= -\int q(x) \log p(x) \mathrm{d}x - H(q) \\ &= c - H(q) - \sum_{v \in \mathcal{V}} \int q_v(x_v) \log \psi_v(x_v) \mathrm{d}x_v \\ &\quad - \sum_{(v_1,v_2) \in \mathcal{E}} \iint q_{(v_1,v_2)}(x_{v_1}, x_{v_2}) \log \psi_{(v_1,v_2)}(x_{v_1}, x_{v_2}) \mathrm{d}x_{v_1} \mathrm{d}x_{v_2} \end{aligned} \tag{27}$$

[13] Other than on a set of zero measure.

where $q_v(\cdot)$ and $q_{(v_1,v_2)}(\cdot,\cdot)$ are the marginal distributions of the random variables $v$ and $(v_1, v_2)$ respectively. This objective can be shown to be equivalent to the *Gibbs free energy*, developed in statistical physics [44].

The only appearance of the full distribution $q(\cdot)$ in (27) is in the entropy term $H(q)$; otherwise only the vertex marginal distributions $q_v(\cdot)$ and the pairwise edge marginals $q_{(v_1,v_2)}(\cdot,\cdot)$ appear. *Variational approximation* methods approach large-scale inference problems for which exact calculations are intractable by approximating the elements of the underlying variational problem, *i.e.*, the objective function and the feasible set [31, Section 4]. In the objective function in (27), this corresponds to replacing the entropy $H(q)$ with an approximation that involves only low order marginal distributions (*e.g.*, vertex and edge marginals). One can then perform the optimisation in (26) directly in terms of these low-order marginals, subject to constraints which ensure that the collection of marginal distributions could feasibly arise from some valid joint distribution.

Many variational approximation methods have been proposed including mean field, belief propagation and tree reweighted sum product [31,45]; here we focus on loopy belief propagation (LBP). A close relationship between LBP and the Bethe free energy[14] was established in [47,48]. Specifically, a problem was studied that was parameterised via vertex and edge marginals, the feasibility constraint was relaxed to a local consistency constraint, *i.e.*,

$$\int \tilde{q}_{(v_1,v_2)}(x_{v_1}, x_{v_2}) \mathrm{d}x_{v_1} = \tilde{q}_{v_2}(x_{v_2}) \ \forall \ (v_1, v_2) \in \mathcal{E}, \ \forall \ x_{v_2} \tag{28}$$

and the objective was replaced with the Bethe free energy:[15]

$$\begin{aligned} \tilde{J}(\tilde{q}) = &- \sum_{v \in \mathcal{V}} \int \tilde{q}_v(x_v) \log \psi_v(x_v) \mathrm{d}x_v \\ &- \sum_{(v_1,v_2) \in \mathcal{E}} \iint \tilde{q}_{(v_1,v_2)}(x_{v_1}, x_{v_2}) \log \psi_{v_1,v_2}(x_{v_1}, x_{v_2}) \mathrm{d}x_{v_1} \mathrm{d}x_{v_2} \\ &- \sum_{v \in \mathcal{V}} H(\tilde{q}_v) + \sum_{(v_1,v_2) \in \mathcal{E}} I(\tilde{q}_{(v_1,v_2)}) \end{aligned} \tag{29}$$

where

$$\begin{aligned} H(\tilde{q}_v) &= - \int \tilde{q}_v(x_v) \log \tilde{q}_v(x_v) \mathrm{d}x_v \\ I(\tilde{q}_{(v_1,v_2)}) &= \iint \tilde{q}_{(v_1,v_2)}(x_{v_1}, x_{v_2}) \log \frac{\tilde{q}_{(v_1,v_2)}(x_{v_1}, x_{v_2})}{\tilde{q}_{v_1}(x_{v_1}) \tilde{q}_{v_2}(x_{v_2})} \mathrm{d}x_{v_1} \mathrm{d}x_{v_2} \end{aligned}$$

The Karush-Kuhn-Tucker (KKT) conditions[16] for this problem provide a series of simultaneous nonlinear equations in the various node and edge marginals.

[14] Named after work by Sir Hans Bethe, Nobel laureate in physics, [46].

[15] We use the equivalent form from [31, p83].

[16] The KKT conditions are necessary conditions for optimality in constrained optimisation problems (and sufficient conditions in convex problems satisfying minor technical conditions) [49].

Manipulating these equations gives rise to the sum-product LBP equations (23), (24), (25); the messages $\mu_{v_1 \to v_2}(\cdot)$ correspond to Lagrange multipliers for the various local consistency constraints [31]. Applying the LBP equations iteratively will attempt to solve the nonlinear equality system using the general iterative method.

Whereas the Gibbs free energy is a convex function of the distribution $q(x)$, the Bethe free energy is generally a non-convex function of the various vertex and edge marginal distributions. This is one of the major causes of the convergence difficulties that LBP experiences in some problems. However, recently it has been shown that in the problem of interest, the Bethe free energy is convex with respect to a particular parameterisation [40,41]; this result leads to a convergence guarantee, and some intuition as to why the marginal probability estimates appear to be of high quality.[17]

## 3.4 Approximating marginal association probabilities

Returning now to our problem of interest, we seek marginal distributions of $a^i$ and $b^j$ where

$$p(a,b) \propto \prod_{i \in \mathcal{T}} \left( w^{i,a^i} \prod_{j \in \mathcal{R}} \psi_r^{i,j}(a^i, b^j) \right) \tag{30}$$

Through (21), these in turn will provide us with Bernoulli multi-target marginal distributions of the various tracks that summarise the distribution of previously detected targets.

First, we restrict ourselves to the case in which all hypotheses $a^i$ consist of either zero or one measurements; this covers the single time step case. The graphical model for this case is shown in Fig. 3. The sum-product messages in this case are: [5]

$$\mu_{a^i \to b^j}(b^j) = \sum_{a^i} w^{i,a^i} \psi_r^{i,j}(a^i, b^j) \prod_{j' \in \mathcal{R} | j' \neq j} \mu_{b^{j'} \to a^i}(a^i)$$
$$\mu_{b^j \to a^i}(a^i) = \sum_{b^j} \psi_r^{i,j}(a^i, b^j) \prod_{i' \in \mathcal{T} | i' \neq i} \mu_{a^{i'} \to b^j}(b^j)$$

### 3.4.1 Simplified message equations

In [36], the MAP version of this problem was studied, and simplified yet equivalent messages were provided. Simplifications for the sum-product case were derived similarly in [5,37]. The simplifications exploit two facts:

- Each matrix $\psi_r^{i,j}(a^i, b^j)$ contains only two unique rows and two unique columns

[17] Empirical evidence suggests that LBP provides good quality estimates when it converges, *e.g.*, [34].

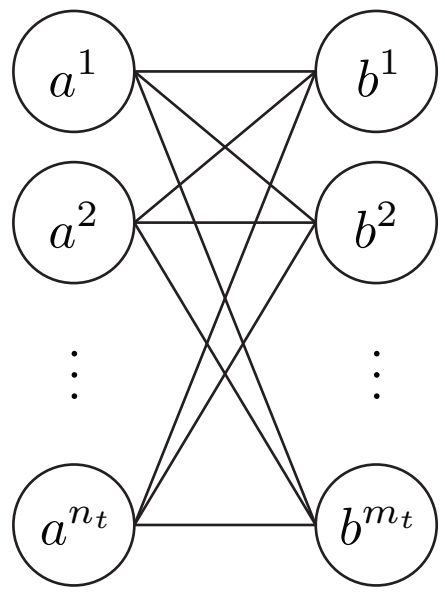

Figure 3: Graphical model depiction of for single scan data association problem. Left-hand variables $a^i$ hypothesise the measurement with which target $i$ is associated; right-hand variables $b^j$ hypothesise the target with which measurement $j$ is associated.

- We are free to renormalise the messages $\mu_{a^i \to b^j}(b^j)$ and $\mu_{b^j \to a^i}(a^i)$

Consequently each message can be parameterised by a *scalar*:

$$\mu_{a^i \to b^j}(b^j) = \begin{cases} \mu_{a^i \to b^j}, & b^j = i \in \mathcal{T} \\ 1, & \text{otherwise} \end{cases}$$

$$\mu_{b^j \to a^i}(a^i) = \begin{cases} \mu_{b^j \to a^i}, & a^i = \{j\} \\ 1, & \text{otherwise} \end{cases}$$

In terms of these scalars, the simplified message updates become:

$$\mu_{a^i \to b^j} = \frac{w^{i,\{j\}}}{w^{i,\emptyset} + \sum_{j' \in \mathcal{R}, j' \neq j} w^{i,\{j'\}} \mu_{b^{j'} \to a^i}} \tag{31}$$

$$\mu_{b^j \to a^i} = \frac{1}{\sum_{i' \in \mathcal{T}, i' \neq i} \mu_{a^{i'} \to b^j}} \tag{32}$$

where we use the convention that $w^{i,a} \triangleq 0$ for $a \notin \mathcal{H}^i$. If the false alarm density is non-zero and the probability of detection is less than unity then the denominators are guaranteed to be non-zero.[18] The non-linear equations (31) and (32) can be iterated until convergence is obtained; Matlab code for a modified version of this process was provided in [5]. Upon convergence, approximate marginals can be recovered as:

$$\tilde{p}^i(a^i) \propto \begin{cases} w^{i,\emptyset}, & a^i = \emptyset \\ w^{i,a^i} \mu_{b^j \to a^i}, & a^i = \{j\}, j \in \mathcal{R} \end{cases} \tag{33}$$

$$\tilde{p}^j(b^j) \propto \mu_{a^i \to b^j}, \quad b^j = i \in \mathcal{T} \tag{34}$$

[18] The only exception to this is the case in which there is a measurement that can only be explained as being a new target; inference in this case is trivial.

### 3.4.2 Comments on accuracy of approximation

The accuracy of this approximation was studied in [4]. The experiments studied problems involving six targets on a regularly spaced grid, where the grid spacing was varied between zero, and a distance such that the targets were non-interacting. The worst case average maximum error in the marginal association weights was 0.006 when $P_d = 0.3$, and 0.083 when $P_d = 0.9$. Holding $P_d = 0.9$ and increasing the false alarm rate also showed a small reduction in the average maximum error. Thus, in the most challenging problems, the approximate weights are closest to the exact values.

Should the accuracy of the approximations provided by this method be insufficient, a range of other methods (*e.g.*, [48], and the emerging result of [41,50]) are available that provide improved accuracy at the expense of increased (but still polynomial) computational complexity.

### 3.4.3 Convergence guarantees and complexity

Convergence of LBP in this problem was proven in parallel in [40,41] and [5]. The proof in [5] assumed a non-unity probability of detection and a non-zero probability of false alarm, and showed that this yields a contraction. The results in [40,41] provide some very interesting properties of the Bethe free energy for the case of interest, and leverages these to establish convergence in the more general case (admitting the case with no false alarms). One result is that the Bethe free energy is convex with respect to a particular parameterisation. In the variational approximation framework, this establishes that LBP is solving a convex approximation to the exact inference problem, in turn providing some intuition as to why the observed empirical performance has been surprisingly good.

## 4 Coalescence avoidance

Using the algorithm proposed in Section 3, targets that are close to each other will compete for measurements, leading to coalescence-like phenomena (*e.g.*, [51]). It is in these same circumstances that convergence of LBP will be at its slowest, and the quality of the approximation in Section 3.4 will be at its poorest. The problem was addressed in [51] by deleting all but the most likely joint hypothesis among groups that are equivalent up to permutation; this requires explicit enumeration of joint (*i.e.*, multi-target) hypotheses. The minimum mean optimal sub-pattern assignment (MMOSPA) filter in [52] computes the estimate that minimises the MOSPA criterion, necessitating a numerical integration in joint target state space. Both of these methods are intractable for all but the simplest problems.

At the other end of the scale, MeMBer avoids coalescence by collecting all hypotheses that were updated with a particular measurement in the latest scan into a single track. While this is computationally tractable, it loses continuity of tracks, and the accuracy of the approximations involved are problematic in low SNR conditions [1, p668, pp678–681].

The proposed method is a hybrid between the previously described RFS filter and a hard, MAP assignment method. Practically, coalescence is one of the major reasons why MAP methods such as MHT are favoured over alternatives such as Gaussian sum filters, so using MAP to resolve coalescence in a marginal track filter is reasonable in principle.

In the single scan case, the problem of finding the MAP hypothesis is easily solvable using the auction algorithm [53]. Subsequently, the weight adjustment can be made by solving the following network flow problem (which can be performed efficiently using the network simplex algorithm [11]). The formulation seeks maximise the transfer of association probability to the MAP association solution. We denote by $\tilde{p}^i(a^i)$ the marginal weight of association event $a^i$ under track $i$ (found approximately via the BPMTF), and by $a^{i*}$ the MAP association event in the 2D assignment ($a^{i*} = \emptyset$ if the target is missed). The network flow formulation is then:

$$\begin{aligned}
\text{maximise} \; & \sum_{i \in \mathcal{T}} \delta^i(a^{i*}) && && (35)\\
\text{subject to} \; & \sum_{a \in \mathcal{H}^i} \delta^i(a) = 0 && \forall\, i \in \mathcal{T} \\
& \sum_{i \in \mathcal{T}} \delta^i(a) = 0 && \forall\, a \in \bigcup_{i \in \mathcal{T}} \mathcal{H}^i \\
& \delta^i(a^{i*}) \geq 0 && \forall\, i \in \mathcal{T} \\
& -\tilde{p}^i(a) q^{i,a} \leq \delta^i(a) \leq 0 && \forall\, i \in \mathcal{T},\; a \in \mathcal{H}^i \backslash \{a^{i*}\} \\
& \delta^i(a) = 0 && \forall\, i \in \mathcal{T},\; a \notin \mathcal{H}^i \\
& \delta^i(\emptyset) = 0 && \forall\, i \in \mathcal{T}
\end{aligned}$$

The final two constraints effectively remove the corresponding variables from the optimisation. Upon completion, the marginal probabilities of every hypothesis of every track are modified to:

$$\begin{aligned}
\bar{p}^i(a) &= \tilde{p}^i(a) + \delta^i(a) \\
\bar{q}^{i,a} &= \frac{\tilde{p}^i(a) q^{i,a} + \delta^i(a)}{\bar{p}^i(a)}
\end{aligned}$$

For each track $i \in \mathcal{T}$ with $a^{i*} \neq \emptyset$, the hypothesis and existence-conditioned distribution is modified to:

$$\bar{f}^{i,a^{i*}}(x) = \frac{\tilde{p}^i(a^{i*}) q^{i,a^{i*}} f^{i,a^{i*}}(x) - \sum_{i' \in \mathcal{T}, i' \neq i} \delta^{i'}(a^{i*}) f^{i',a^{i*}}(x)}{\bar{p}^i(a^{i*}) \bar{q}^{i,a^{i*}}}$$

For remaining hypotheses, $\bar{f}^{i,a}(x) = f^{i,a}(x)$. Appendix B shows that the update rules modify neither the existence probability of each track nor the overall first moment of the distribution.

This exchange results in a loss of track identity information; the resulting distribution of track identity can be maintained through an additional layer (*e.g.*, [54]).

# 5 BP-MeMBer filter

The previous sections have described the BPMTF, and the necessary modifications for coalescence avoidance. As described in Section 2.5, one key difference between the proposed method and the MeMBer is the reversal of the hypothesis branching structure. This section describes an algorithm which we term the belief propagation multi-target multi-Bernoulli filter (BP-MeMBer), which forms tracks in a similar manner to the MeMBer, but uses the LBP methods described in Section 3 to calculate marginal probabilities that are cognisant of the constraints from the previous time step.

Consider the information contained in the two sets of association variables, $a = (a^i)_{i\in\mathcal{T}}$ and $b = (b^j)_{j\in\mathcal{R}}$. The value of the random variable $a^i$ is the measurement (subset) associated with target $i$, while the value of $b^j$ is the index of the target associated with measurement $j$. These two sets are redundant: given $a$ one can perfectly reconstruct $b$, and vice versa. Accordingly, we can represent the updated multi-target PGFl equivalently as

$$G^p_{t+1|t+1}[h] = \sum_a p_{t+1|t+1}(a) \prod_{i\in\mathcal{T}_{t+1}} G^{i,a^i}_{t+1|t+1}[h]$$

or

$$G^p_{t+1|t+1}[h] = \sum_b p_{t+1|t+1}(b) \prod_{i\in\mathcal{T}_{t+1}} G^{i,\{j\in\mathcal{R}_{t+1}|b^j=i\}}_{t+1|t+1}[h]$$

The essential difference between the BPMTF and the BP-MeMBer is that of approximating the target association distribution as $p(a) \approx \prod_{i\in\mathcal{T}} p^i(a^i)$ (in the BPMTF), versus approximating the equivalent measurement association distribution as $p(b) \approx \prod_{j\in\mathcal{R}} p^j(b^j)$ (in the BP-MeMBer).

To proceed, assume that the predicted distribution carried forward from the previous time step is multi-Bernoulli, *i.e.*, there is a set of tracks $\mathcal{T}_t$, each of which consists of a single hypothesis $\mathcal{H}^i_{t+1|t} = \{\emptyset\}$.[19] Let $n_t = |\mathcal{T}_t|$, and denote by $Z_{t+1} = \{z^1, \dots, z^{m_{t+1}}\}$ the set of measurements at time $(t+1)$, so that $\mathcal{T}_{t+1} = \{1, \dots, n_t + m_{t+1}\}$. Let $\mathcal{R}_{t+1} = \{1, \dots, m_{t+1} + n_t\}$; resources $\{1, \dots, m_{t+1}\}$ correspond to the measurements, while resources $\{m_{t+1} + 1, \dots, m_{t+1} + n_t\}$ are pseudo-measurements corresponding to the respective tracks $\{1, \dots, n_t\}$ experiencing a missed detection. Accordingly, each input track $i$ will result in $(m_{t+1} + n_t)$ updated hypotheses, with $(n_t - 1)$ of these having zero weight (since $w^{i,m_{t+1}+j}_{t+1|t+1} = 0$, $j \neq i$, $j > 0$ by construction), and similarly each new target track $i \in \{n_t + 1, \ldots, n_t + m_{t+1}\}$ will consist of $(n_t + m_{t+1})$ hypotheses where only a single hypothesis has non-zero weight. We can exactly represent the updated PGFl as:

$$G^p_{t+1|t+1}[h] = \sum_b p_{t+1|t+1}(b) \prod_{j \in \mathcal{R}_{t+1}} G^{b^j,\{j\}}_{t+1|t+1}[h]$$

[19]This single hypothesis may involve a multi-modal distribution, but there are no association constraints between targets. The hypothesis is denoted as $\emptyset$ association constraints from previous scans are represented by adjusting the conditional track distribution using the marginal association probabilities.

| | | Track ($i \in \mathcal{T}$) | | | | |
|---|---|---|---|---|---|---|
| | | Existing tracks | | | New tracks | |
| Measurement ($j \in \mathcal{R}$) | | 1 | 2 | 3 | 4 | 5 |
| Measurements | 1 | 0.5 | 0.2 | 0.1 | 0.2 | 0 |
| | 2 | 0.1 | 0.2 | 0.6 | 0 | 0.1 |
| Pseudo-measurements | 3 | 0.4 | 0 | 0 | 0 | 0 |
| (missed detections) | 4 | 0 | 0.6 | 0 | 0 | 0 |
| | 5 | 0 | 0 | 0.3 | 0 | 0 |

Table 1: Example case showing difference between BPMTF and BP-MeMBer. Tracks carried forward in BPMTF correspond to columns of table, whereas tracks in BP-MeMBer correspond to rows in table.

Applying the alternative approximation $p_{t+1|t+1}(b) \approx \prod_{j \in \mathcal{R}_{t+1}} p^j_{t+1|t+1}(b^j)$, we obtain

$$G^p_{t+1|t+1}[h] \approx \prod_{j \in \mathcal{R}_{t+1}} \sum_{b^j} p^j_{t+1|t+1}(b^j) G^{b^j,\{j\}}_{t+1|t+1}[h]$$

The BP-MeMBer uses this expression, replacing the marginal probability $p^j_{t+1|t+1}(b^j)$ with its LBP approximation $\tilde{p}^j_{t+1|t+1}(b^j)$.

To provide a concrete example, consider a case in which the marginal probabilities are as detailed in Table 1. Using the BPMTF, one track would be generated for each column in the table. Conversely, using the BP-MeMBer, one track would be generated for each row in the table. The two methods provide alternative approximations of the joint distribution.

# 6 Undetected targets and recycling

Recall again the form that the BPMTF maintains from one time interval to the next:

$$G_{t|t}[h] \propto \exp\{\lambda^u_{t|t}[h]\} \cdot \prod_{i \in \mathcal{T}_t} G^i_{t|t}[h] \tag{36}$$

Note that this is a union of independent components: a Poisson point process component, and a series of independent multi-target Bernoulli components.[20] Since a new track must be created for every new measurement received, tracks will inevitably need to be deleted from the system. The system designer must trade off system performance against computational complexity in constructing this mechanism.

[20] The multi-target Bernoulli components are not naturally independent but they are *approximated* as such, as in M-C [2], and the MeMBer [1].

One possibility which arises from the form of (36) is to approximate individual tracks as being Poisson, rather than deleting them. Consider a single multi-target Bernoulli component:

$$f^i_{t|t}(X) = \begin{cases} 1 - q^i_{t|t}, & X = \emptyset \\ q^i_{t|t} f^i_{t|t}(x), & X = \{x\} \\ 0, & |X| > 1 \end{cases}$$

We may choose to approximate this component as being Poisson:

$$\tilde{f}^i_{t|t}(X) = \exp(-\lambda^i_{t|t}) \prod_{x \in X} \lambda^i_{t|t}(x)$$

where $\lambda^i_{t|t} = \int \lambda^i_{t|t}(x)\mathrm{d}x$, and we define $\tilde{f}^i_{t|t}(x) \triangleq \lambda^i_{t|t}(x)/\lambda^i_{t|t}$ for later use. The distortion caused by this approximation may be measured by the multi-target KL divergence: [1, p513]

$$\begin{aligned} D(f^i_{t|t}(X)||\tilde{f}^i_{t|t}(X)) &= \int f^i_{t|t}(X) \log \frac{f^i_{t|t}(X)}{\tilde{f}^i_{t|t}(X)} \delta X \\ &= (1 - q^i_{t|t}) \log \frac{1 - q^i_{t|t}}{\exp(-\lambda^i_{t|t})} + \int q^i_{t|t} f^i_{t|t}(x) \log \frac{q^i_{t|t} f^i_{t|t}(x)}{\exp(-\lambda^i_{t|t}) \lambda^i_{t|t} \tilde{f}^i_{t|t}(x)} \mathrm{d}x \\ &= (1 - q^i_{t|t}) \log(1 - q^i_{t|t}) + (1 - q^i_{t|t}) \lambda^i_{t|t} + q^i_{t|t} [\log q^i_{t|t} + \lambda^i_{t|t} - \log \lambda^i_{t|t}] \\ &\quad + q^i_{t|t} D(f^i_{t|t}(x)||\tilde{f}^i_{t|t}(x)) \end{aligned}$$

It is optimal to set $\tilde{f}^i_{t|t}(x) = f^i_{t|t}(x)$ and $\lambda^i_{t|t} = q^i_{t|t}$, so that $\lambda^i_{t|t}(x) = q^i_{t|t} f^i_{t|t}(x)$ (as shown in [1, p579]). The value of the KL divergence at this optimal choice is:

$$D(f^i_{t|t}(X)||\tilde{f}^i_{t|t}(X)) = q^i_{t|t} + (1 - q^i_{t|t}) \log(1 - q^i_{t|t})$$

The value of this KL divergence is shown in Fig. 4 as a function of the target existence $q^i_{t|t}$. The figure demonstrates that the distortion caused by the approximation (as measured by the KL divergence) is very small for existence probabilities less than 0.1.

In Appendix C we show that the KL divergence between the overall multi-target distribution comprised of independent components (*e.g.*, (36)) and a modified multi-target distribution in which approximations have been made to a number of these components is bounded by the sum of the KL divergences between the components and their respective approximations. Thus the overall distortion we cause to the complete multi-target distribution is bounded by the sum of the individual track distortions, which depends only on the track existence probability. Accordingly, we may approximate the tracks with lowest existence probabilities such that the sum of the distortions is less than an overall distortion budget.

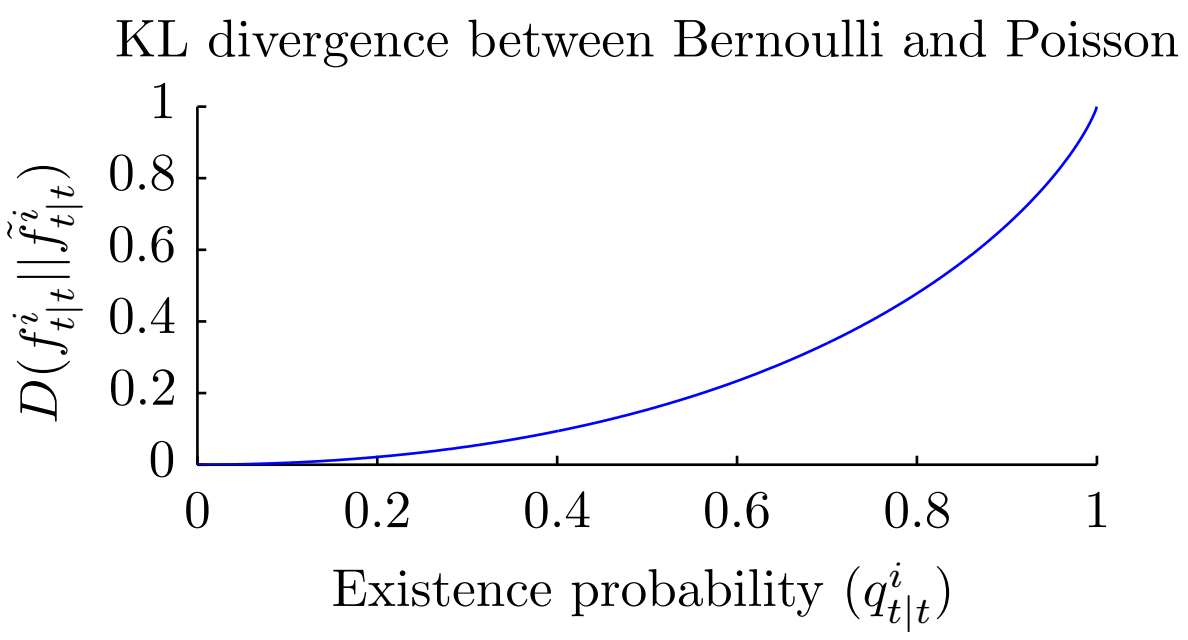

Figure 4: Multi-target KL divergence between multi-target Bernoulli distribution and best-fit Poisson distribution as a function of target existence probability $q^i_{t|t}$.

When any number of Bernoulli tracks are approximated as being Poisson, the resulting multi-target distribution is equivalent to one in which those tracks are dropped, and their intensity is added onto the undetected target distribution. To confirm this, denote the subset of tracks that we retain as $\tilde{\mathcal{T}}_t$; the approximated distribution is then

$$\begin{aligned}
\tilde{G}_{t|t}[h] &\propto \exp\{\lambda^u_{t|t}[h]\} \cdot \prod_{i\in\mathcal{T}_t\setminus\tilde{\mathcal{T}}_t} \tilde{G}^i_{t|t}[h] \cdot \prod_{i\in\tilde{\mathcal{T}}_t} G^i_{t|t}[h] \\
&\propto \exp\{\lambda^u_{t|t}[h]\} \cdot \prod_{i\in\mathcal{T}_t\setminus\tilde{\mathcal{T}}_t} \exp\{\lambda^i_{t|t}[h]\} \cdot \prod_{i\in\tilde{\mathcal{T}}_t} G^i_{t|t}[h] \\
&= \exp\left\{\lambda^u_{t|t}[h] + \sum_{i\in\mathcal{T}_t\setminus\tilde{\mathcal{T}}_t} \lambda^i_{t|t}[h]\right\} \cdot \prod_{i\in\tilde{\mathcal{T}}_t} G^i_{t|t}[h] \\
&= \exp\{\tilde{\lambda}^u_{t|t}[h]\} \cdot \prod_{i\in\tilde{\mathcal{T}}_t} G^i_{t|t}[h]
\end{aligned}$$

where

$$\tilde{\lambda}^u_{t|t}[h] \triangleq \lambda^u_{t|t}[h] + \sum_{i\in\mathcal{T}_t\setminus\tilde{\mathcal{T}}_t} \lambda^i_{t|t}[h]$$

We refer to this concept as *recycling*, since the tracks that we delete re-enter the system through the undetected target intensity. Not surprisingly, it can be shown that if the prior distribution is purely Poisson (*i.e.*, there are no pre-existing target tracks) and we choose to recycle all posterior tracks, then the posterior Poisson distribution is equivalent to that obtained using the PHD. By recycling a subset of tracks, we permit the large mass of tracks with low probability of existence to be represented efficiently by the Poisson distribution, while maintaining explicit Bernoulli tracks on the subset with non-negligible probability of detection. Representing low probability of existence tracks via the Poisson distribution reduces the computational complexity due to data association. Furthermore, if the Poisson distribution is represented as a discrete grid, then there is no computational cost associated with representing additional tracks.

In practice, this approximation allows the system to gradually accrue confidence in the presence of a target before choosing to maintain an explicit track. As shown in equation (13), the existence probability is proportional to the undetected target intensity (in the vicinity of the measurement) divided by the undetected target intensity plus the false alarm intensity, so the intensity added by recycling will cause the existence probability of a new track due to a later measurement in the same vicinity to be increased, reducing the likelihood that the track will again be recycled. This structure is likely to be most advantageous in very low SNR environments, where there is a large number of false alarms, and confidence in target existence must be accrued over a significant time.

# 7 Conclusion

We have shown that, under common assumptions, the full RFS distribution can be represented via a structured form, which admits application of graphical model inference methods. A heuristic, computationally tractable method was presented for addressing the problem of coalescence in multiple target tracking problems. We described the benefit of maintaining a discretised Poisson representation of undetected targets alongside the conventional representation, introducing the concept of target recycling, which maintains the first moment of the distribution upon deletion of tracks. The next step in this research is to compare the proposed methods to existing filters such as the [C]PHD and MeMBer.

# A Linearity of probability generating functional

In general, linearity of a probability generating functional is straight-forward, following directly from linearity of the integration operator. This section confirms that this remains the case in the random finite set case, in which the integral is a set integral. The required property is that:

**Lemma 2.** *The RFS probability generating functional is linear, i.e.,*

$$\int h^X[\alpha f_1(X) + \beta f_2(X)]\delta X = \alpha \int h^X f_1(X)\delta X + \beta \int h^X f_2(X)\delta X$$

*Proof.* This follows simply since:

$$
\begin{aligned}
&\int h^X[\alpha f_1(X)+\beta f_2(X)]\delta X \\
&\quad = [\alpha f_1(\emptyset)+\beta f_2(\emptyset)]+ \\
&\qquad + \sum_{n=1}^{\infty}\int\left[\prod_{i=1}^{n} h(x_i)\right]\cdot[\alpha f_1(\{x_1,\ldots,x_n\})+\beta f_2(\{x_1,\ldots,x_n\})]\,\mathrm{d}x_1,\cdots,\mathrm{d}x_n \\
&\quad = \alpha\left\{f_1(\emptyset)+\sum_{n=1}^{\infty}\int\left[\prod_{i=1}^{n} h(x_i)\right] f_1(\{x_1,\ldots,x_n\})\mathrm{d}x_1,\cdots,\mathrm{d}x_n\right\}+ \\
&\qquad + \beta\left\{f_2(\emptyset)+\sum_{n=1}^{\infty}\int\left[\prod_{i=1}^{n} h(x_i)\right] f_2(\{x_1,\ldots,x_n\})\mathrm{d}x_1,\cdots,\mathrm{d}x_n\right\} \\
&\quad = \alpha\int h^X f_1(X)\delta X+\beta\int h^X f_2(X)\delta X
\end{aligned}
$$

□

Note that this is indeed a general property of set integrals, applying not only to the PGFl transformation.

# B Properties of coalescence avoidance algorithm

This section establishes two properties that are maintained by the coalescence avoidance algorithm in Section 4. First, we show that the probability of existence of each track is unmodified.

**Lemma 3.** *For each track $i \in \mathcal{T}$,*

$$\bar{q}^i \triangleq \sum_{a\in\mathcal{H}^i} \bar{p}^i(a)\bar{q}^{i,a} = \sum_{a\in\mathcal{H}^i} \tilde{p}^i(a)q^{i,a} \triangleq q^i$$

*Proof.*

$$\sum_{a\in\mathcal{H}^i} \bar{p}^i(a)\bar{q}^{i,a} = \sum_{a\in\mathcal{H}^i} [\tilde{p}^i(a)q^{i,a}+\delta^i(a)] = \sum_{a\in\mathcal{H}^i} \tilde{p}^i(a)q^{i,a}$$

since $\sum_{a\in\mathcal{H}^i}\delta^i(a)=0$. □

**Lemma 4.** *The network flow coalescence updates in Section 4 do not change the first moment of the distribution, i.e.,*

$$\sum_{i\in\mathcal{T}}\sum_{a\in\mathcal{H}^i} \bar{p}^i(a)\bar{q}^{i,a}\bar{f}^{i,a}(x) = \sum_{i\in\mathcal{T}}\sum_{a\in\mathcal{H}^i} \tilde{p}^i(a)q^{i,a}f^{i,a}(x)$$

*Proof.*

$$
\begin{aligned}
D(x) &= \sum_{i \in \mathcal{T}} \sum_{a \in \mathcal{H}^i} \bar{p}^i(a) \bar{q}^{i,a} \bar{f}^{i,a}(x) \\
&= \sum_{i \in \mathcal{T}} \left[ \bar{p}^i(a^{i*}) \bar{q}^{i,a^{i*}} \bar{f}^{i,a^{i*}}(x) + \sum_{a \in \mathcal{H}^i \backslash \{a^{i*}\}} \bar{p}^i(a) \bar{q}^{i,a} f^{i,a}(x) \right] \\
&= \sum_{i \in \mathcal{T}} \left[ \tilde{p}^i(a^{i*}) q^{i,a^{i*}} f^{i,a^{i*}}(x) - \sum_{i' \in \mathcal{T} | i' \neq i} \delta^{i'}(a^{i*}) f^{i',a^{i*}}(x) + \right. \\
&\qquad \left. + \sum_{a \in \mathcal{H}^i \backslash \{a^{i*}\}} [\tilde{p}^i(a) q^{i,a} + \delta^i(a)] f^{i,a}(x) \right] \\
&= \sum_{i \in \mathcal{T}} \sum_{a \in \mathcal{H}^i} \tilde{p}^i(a) q^{i,a} f^{i,a}(x) + \\
&\qquad + \sum_{i \in \mathcal{T}} \left[ \sum_{a \in \mathcal{H}^i \backslash \{a^{i*}\}} \delta^i(a) f^{i,a}(x) - \sum_{i' \in \mathcal{T} \backslash \{i\}} \delta^i(a^{i'*}) f^{i,a^{i'*}}(x) \right]
\end{aligned}
$$

Consider a term in the final line, $i \in \mathcal{T}$ and $a \in \mathcal{H}^i \backslash \{a^{i*}\}$. Suppose that $\nexists\ i'$ such that $a^{i'*} = a$. Then $\delta^i(a) = 0\ \forall\ i$ since $\delta^i(a) \leq 0\ \forall\ i$ and $\sum_i \delta^i(a) = 0$. Now suppose that there is a unique $i'$ with $a^{i'*} = a$. In this case, the negative term in the second sum on the final line will cancel the term in the first sum. Finally, suppose that there are multiple $i'$ with $a^{i'*} = a$. By the constraints on the MAP solution, the only way this can happen is if $a = \emptyset$, so that $\delta^i(a) = 0$. Thus the final line is zero. □

# C Sub-additivity of KL divergence

The following theorem establishes a result that the KL divergence between a multi-target distribution and the distribution that results from approximations made to a series of independent subcomponents of the distribution is bounded by the sum of the KL divergences between the subcomponents and their respective approximations. We prove the result for two independent subcomponents; the general case results simply from induction.

**Theorem 1.** *Let*

$$f(X) = \sum_{W \subseteq X} g(W) h(X - W)$$

*and*

$$\tilde{f}(X) = \sum_{W \subseteq X} \tilde{g}(W) \tilde{h}(X - W)$$

*Then*

$$D(f || \tilde{f}) \leq D(g || \tilde{g}) + D(h || \tilde{h})$$

*Proof.*

$$D(f||\tilde{f}) = \int \left[\sum_{W\subseteq X} g(W)h(X-W)\right] \log \frac{\left[\sum_{W\subseteq X} g(W)h(X-W)\right]}{\left[\sum_{W\subseteq X} \tilde{g}(W)\tilde{h}(X-W)\right]} \delta X$$

$$\stackrel{(a)}{\leq} \int \sum_{W\subseteq X} g(W)h(X-W) \log \frac{g(W)h(X-W)}{\tilde{g}(W)\tilde{h}(X-W)} \delta X$$

$$= \int \sum_{W\subseteq X} g(W)h(X-W) \log \frac{g(W)}{\tilde{g}(W)} \delta X$$

$$+ \int \sum_{W\subseteq X} g(W)h(X-W) \log \frac{h(X-W)}{\tilde{h}(X-W)} \delta X$$

$$\stackrel{(b)}{=} D(g||\tilde{g}) + D(h||\tilde{h})$$

where $(a)$ is a consequence of the log-sum inequality, [42, p29], and $(b)$ is the result of Lemma 5. □

**Lemma 5.**

$$\int \sum_{W\subseteq X} a(W)b(X-W)\delta X = \int a(W)\delta W \cdot \int b(Y)\delta Y$$

*Proof.* For simplicity, let $\int \cdots \int a(\{x_1,\ldots,x_n\})\mathrm{d}x_1\cdots\mathrm{d}x_n \triangleq a(\emptyset)$ when $n=0$. Then:

$$\int \sum_{W\subseteq X} a(W)b(X-W)\delta X$$

$$\triangleq \sum_{n=0}^{\infty} \frac{1}{n!} \int \cdots \int \sum_{W\subseteq\{x_1,\ldots,x_n\}} a(W)b(\{x_1,\ldots,x_n\}-W)\mathrm{d}x_1\cdots\mathrm{d}x_n$$

$$= \sum_{n=0}^{\infty} \sum_{I\subseteq\{1,\ldots,n\}} \frac{1}{n!} \int \cdots \int a(\{x_i|i\in I\})b(\{x_i|i\notin I\})\mathrm{d}x_1\cdots\mathrm{d}x_n$$

$$\stackrel{(a)}{=} \sum_{n=0}^{\infty} \sum_{m=0}^{n} \frac{1}{n!} \cdot \frac{n!}{m!(n-m)!} \cdot \int \cdots \int a(\{x_1,\ldots,x_m\})\mathrm{d}x_1\cdots\mathrm{d}x_m \cdot$$

$$\cdot \int \cdots \int b(\{x_{m+1},\ldots,x_n\})\mathrm{d}x_{m+1}\cdots\mathrm{d}x_n$$

$$\stackrel{(b)}{=} \sum_{m=0}^{\infty} \sum_{l=0}^{\infty} \frac{1}{m!} \int \cdots \int a(\{x_1,\ldots,x_m\})\mathrm{d}x_1\cdots\mathrm{d}x_m \cdot$$

$$\cdot \frac{1}{l!} \int \cdots \int b(\{x_1,\ldots,x_l\})\mathrm{d}x_1\cdots\mathrm{d}x_l$$

$$= \int a(W)\delta W \cdot \int b(Y)\delta Y$$

where $(a)$ results from the observation that terms in the sum over $I$ with the same number of elements in $I$ will have the same value, and $(b)$ results from letting $l = (n - m)$ and reordering terms in the two-dimensional summation. $\square$